\documentclass[a4paper]{jpconf}
\usepackage{graphicx}
\begin{document}
\title{KLEVER: An experiment to measure
  \boldmath{${\rm BR}(K_L\to\pi^0\nu\bar{\nu})$} at the CERN SPS}

\author{M Moulson, for the KLEVER Project\footnote{
The KLEVER Project: 
F~Ambrosino,
R~Ammendola,
A~Antonelli,
K~Ayers,
D~Badoni,
G~Ballerini,
L~Bandiera,
J~Bernhard,
C~Biino,
L~Bomben,
V~Bonaiuto,
A~Bradley,
M~B~Brunetti,
F~Bucci,
A~Cassese,
R~Camattari,
M~Corvino,
D~De~Salvador,
D~Di~Filippo,
M~van~Dijk,
N~Doble,
R~Fantechi,
S~Fedotov,
A~Filippi,
F~Fontana,
L~Gatignon,
G~Georgiev,
A~Gerbershagen,
A~Gianoli,
E~Imbergamo,
K~Kampf,
M~Khabibullin,
S~Kholodenko,
A~Khotjantsev,
V~Kozhuharov,
Y~Kudenko,
V~Kurochka,
G~Lamanna,
M~Lenti,
L~Litov,
E~Lutsenko,
T~Maiolino,
I~Mannelli,
S~Martellotti,
M~Martini,
V~Mascagna,
A~Maslenkina,
P~Massarotti,
A~Mazzolari,
E~Menichetti,
O~Mineev,
M~Mirra,
M~Moulson,
I~Neri,
M~Napolitano,
V~Obraztsov,
A~Ostankov,
G~Paoluzzi,
F~Petrucci,
M~Prest,
M~Romagnoni,
M~Rosenthal,
P~Rubin,
A~Salamon,
G~Salina,
F~Sargeni,
A~Shaykhiev,
A~Smirnov,
M~Soldani,
D~Soldi,
M~Sozzi,
V~Sugoniaev,
A~Sytov,
E~Vallazza,
R~Volpe,
R~Wanke,
N~Yershov}
  }

\address{INFN Laboratori Nazionali di Frascati, 00044 Frascati RM, Italy}

\ead{moulson@lnf.infn.it}

\begin{abstract}
  Precise measurements of the branching ratios for the flavor-changing
  neutral current decays $K\to\pi\nu\bar{\nu}$ can provide unique
  constraints on CKM unitarity and, potentially, evidence for new physics.
  It is important to measure both decay modes, $K^+\to\pi^+\nu\bar{\nu}$ and
  $K_L\to\pi^0\nu\bar{\nu}$, since different new physics models affect
  the rates for each channel differently.
  The NA62 experiment at the CERN SPS will measure the BR for the charged
  channel to better than 20\%.
  The BR for the neutral channel has never been measured.
  We are designing the KLEVER experiment to measure
  BR($K_L\to\pi^0\nu\bar{\nu}$) to $\sim$20\% using a high-energy
  neutral beam at the CERN SPS.
  The boost from the high-energy beam facilitates the rejection of background
  channels such as $K_L\to\pi^0\pi^0$ by detection of the additional photons
  in the final state.
  On the other hand, the layout poses particular challenges for the
  design of the small-angle vetoes, which must reject photons from $K_L$
  decays escaping through the beam exit amid an intense background from
  soft photons and neutrons in the beam.
  We present findings from our design studies,
  with an emphasis on the challenges faced and the
  potential sensitivity for the measurement of
  BR($K_L\to\pi^0\nu\bar{\nu}$).
  \end{abstract}

\section{Introduction}

The branching ratios (BRs) for the decays $K\to\pi\nu\bar{\nu}$
are among the observables in the quark flavor sector most sensitive to
new physics. The Standard Model (SM) rates for these flavor-changing
neutral-current decays are very strongly suppressed by the GIM mechanism
and the CKM hierarchy.
In addition, because of the dominance of the diagrams with top loops,
the lack of contributions from intermediate photons, and the fact that 
the hadronic matrix element can be obtained from $K_{e3}$ data,
the SM rates can be
calculated very precisely: 
${\rm BR}(K^+\to\pi^+\nu\bar{\nu}) = (8.4\pm1.0)\times10^{-11}$ and 
${\rm BR}(K_L\to\pi^0\nu\bar{\nu}) = (3.4\pm0.6)\times10^{-11}$, where
the uncertainties are dominated by the external contributions from
$V_{cb}$ and $V_{ub}$ and the non-parametric theoretical uncertainties
are about 3.5\% and 1.5\%, respectively~\cite{Buras:2015qea}.

Because these decays are strongly suppressed and calculated very
precisely in the SM, their BRs are
potentially sensitive to
mass scales of hundreds of TeV, surpassing the sensitivity
of $B$ decays in most SM extensions~\cite{Buras:2014zga}.
Observations of lepton-flavor-universality-violating phenomena are mounting
in the $B$ sector~\cite{Bifani:2018zmi}.
Most explanations for such phenomena
predict strong third-generation couplings and thus significant changes
to the $K\to\pi\nu\bar{\nu}$ BRs through couplings to final states with
$\tau$ neutrinos~\cite{Bordone:2017lsy}.
Measurements of the $K\to\pi\nu\bar{\nu}$ BRs are critical
to interpreting the data from rare $B$ decays, and may demonstrate that
these effects are a manifestation of new degrees of freedom such as
leptoquarks~\cite{Buttazzo:2017ixm,Fajfer:2018bfj}.
Because the amplitude for $K^+\to\pi^+\nu\bar{\nu}$
has both real and imaginary parts, while the amplitude for
$K_L\to\pi^0\nu\bar{\nu}$ is
purely imaginary, the two decays have different sensitivity to new
sources of $CP$ violation.
Measurements of both BRs would therefore be extremely useful not only
to uncover evidence of new physics in the quark-flavor sector, but also
to distinguish between new physics models.

The BR for the charged decay $K^+\to\pi^+\nu\bar{\nu}$ was first
measured by Brook\-ha\-ven experiment E787 and its successor, E949,
using $K^+$ decays at rest \cite{Artamonov:2009sz}.
The NA62 experiment at the CERN SPS has recently obtained a stronger
constraint on the BR using decay-in-flight data collected in 2016--2017.
The preliminary NA62 result is
${\rm BR}(K^+\to\pi^+\nu\bar{\nu}) < 24.4\times10^{-11}$ (95\% CL),
with a single-event sensitivity (SES) of $(3.46\pm0.17)\times10^{-11}$,
$1.65\pm0.31$ expected background events, and three observed signal
candidates \cite{Ruggiero}.
Additional running is contemplated beginning in 2021, with the goal of
measuring the BR to 20\% precision or better~\cite{NA62:2019xxx}.

The decay $K_L\to\pi^0\nu\bar{\nu}$ has never been measured,
and the KOTO experiment at J-PARC is the only experiment currently
pursuing its observation.
The salient features of the KOTO experiment are the use of a highly collimated,
low-energy (mean momentum 2.1~GeV) ``pencil'' beam and very-high-performance
hermetic calorimetry and photon vetoing.
From an analysis of 2015 data, KOTO has recently obtained the limit
${\rm BR}(K_L\to\pi^0\nu\bar{\nu}) < 3.0\times10^{-9}$ (90\% CL),
with an expected background of $0.42\pm0.18$ events and no candidate
events observed~\cite{Ahn:2019xxx}.
More recently, in a preliminary analysis of data collected in 2016--2018,
the experiment has achieved an SES of $6.9\times10^{-10}$ with
$0.05\pm0.02$ background events expected. Four events found in the signal
region are under further investigation~\cite{Shinohara}.
KOTO should reach an SES equal to the SM BR for the decay by the mid 2020s,
and upgrades to the experiment planned or in progress should reduce
backgrounds by a similar factor~\cite{Tung}.
This would allow a 90\% CL upper limit on the BR to be set at the $10^{-10}$
level, but a next-generation experiment is needed in order to actually
measure the BR.

\section{KLEVER}

We are designing the KLEVER experiment to use a high-energy neutral beam
at the CERN SPS to achieve a sensitivity of about 60 events for the decay
$K_L\to\pi^0\nu\bar{\nu}$ at the SM BR with an $S/B$ ratio of 1.
At the SM BR, this would correspond to a relative uncertainty of about 20\%.
We would expect to observe a discrepancy with $5\sigma$ significance
if the true rate is a bit more than twice or less than one-quarter of the
SM rate, or with $3\sigma$ significance if the true rate is less than half
of the SM rate.
These scenarios are consistent with the rates predicted for many different
SM extensions~\cite{Ambrosino:2019xxx}.

\begin{figure}[htb]
  \centering
  \includegraphics[width=0.8\textwidth]{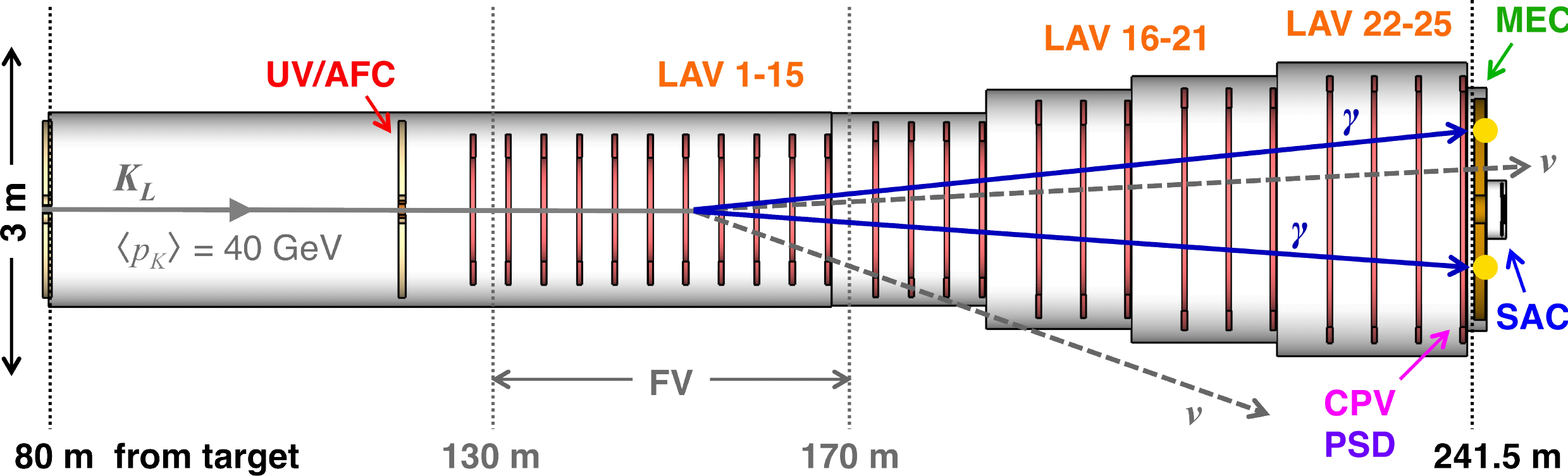}
  \caption{KLEVER experimental apparatus: upstream veto (UV) and
    active final collimator (AFC), large-angle photon vetoes (LAV), main
    electromagnetic calorimeter (MEC), small-angle calorimeter (SAC),
    charged particle veto (CPV), preshower detector (PSD).}
  \label{fig:exp}
\end{figure}
The KLEVER timescale is driven by the assumption that the multi-decade
North Area program in kaon physics, building on the successes of NA62
and its progenitors, will have advanced by the
end of LHC Run 3 to the point at which the measurement of
$K_L\to\pi^0\nu\bar{\nu}$ is the natural next step.
KLEVER would aim to start data taking in LHC Run 4 (2026).
The layout is sketched in Figure~\ref{fig:exp}. Relative to KOTO,
the boost from the high-energy beam in KLEVER facilitates the rejection
of background channels such as $K_L\to\pi^0\pi^0$ by detection of the
additional photons in the final state.
On the other hand, the layout poses particular challenges for the
design of the small-angle vetoes, which must reject photons from $K_L$
decays escaping through the beam exit amidst an intense background from
soft photons and neutrons in the beam. Background from $\Lambda \to n\pi^0$
decays in the beam must also be kept under control.

\subsection{Beam}

KLEVER will make use of the 400-GeV SPS proton beam to produce a neutral
secondary beam at an angle of 8 mrad with an opening angle of 0.4 mrad.
The choice of production angle balances increased $K_L$ production and higher
$K_L$ momentum at small angle with improved $K_L/n$ and $K_L/\Lambda$ ratios
at larger angles. The choice of solid angle balances beam flux against
the need for tight collimation for increased $p_\perp$ constraints to reject
the main background from $K_L\to\pi^0\pi^0$ with lost
photons. The resulting neutral beam has
a mean $K_L$ momentum of 40 GeV, leading to an acceptance of
4\% for the fiducial volume extending from 130 to 170 m downstream of the
target, and a $K_L$ yield of $2\times10^{-5}$ $K_L$ per proton on target (pot).
With a selection
efficiency of 5\%, collection of 60 SM events would require a total
primary flux of $5\times10^{19}$ pot, corresponding to an intensity of
$2\times10^{13}$ protons per pulse (ppp) under NA62-like slow-extraction
conditions, with
a 16.8~s spill cycle and 100 effective days of running per year.
This is a six-fold increase in the primary intensity relative to NA62.
The feasibility of upgrades to the P42 beamline, TCC8 target gallery,
and ECN3 experimental cavern to handle this intensity has been studied
by the Conventional Beams working group in the context of the 
Physics Beyond Colliders initiative, and preliminary
indications are positive~\cite{Banerjee:2018xxx,Gatignon:2018xxx}.

\begin{figure}[htb]
  \centering
  \includegraphics[width=0.7\textwidth]{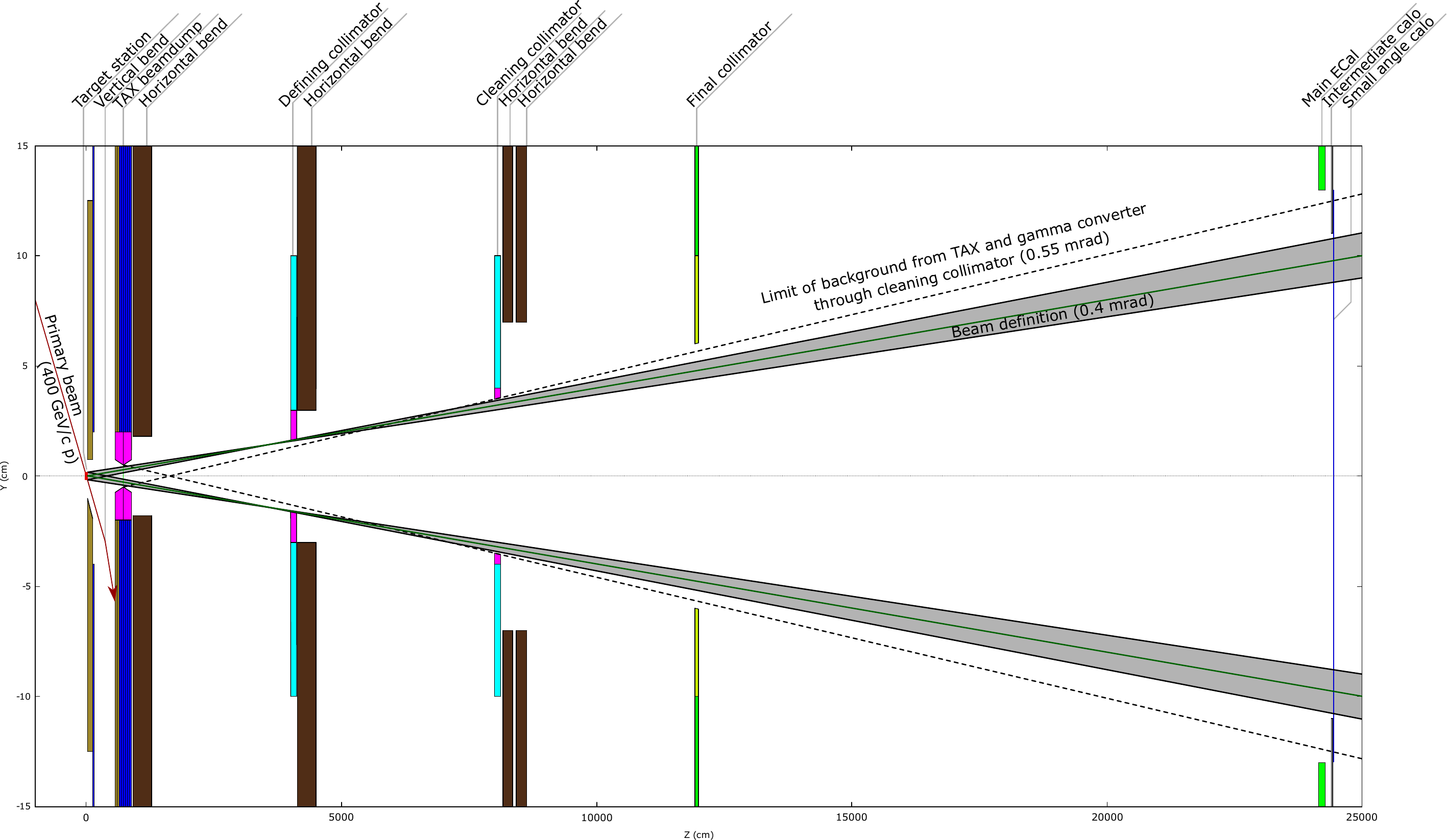}
  \caption{KLEVER neutral beamline layout as modeled in FLUKA,
    showing four collimation stages
    corresponding to dump, defining, cleaning, and final collimators.
    The final collimator is an active detector (AFC), built into the
    upstream veto. The aperture of the main calorimeter and coverage of the
    small-angle calorimeters is also shown.}
  \label{fig:beam}
\end{figure}

A four-collimator neutral beamline layout for ECN3 has been developed, as
illustrated in Fig.~\ref{fig:beam}. The primary proton beam, sloping
downwards at an angle of 8 mrad, is incident on the T10 target (assumed
to be a beryllium rod of 2~mm radius), producing the neutral
beam and other secondaries. This is immediately followed by a vertical
sweeping magnet that bends the protons further downward by 5.6 mrad,
and a TAX dump collimator with a hole for the passage of the neutral beam.
A horizontal sweeping magnet downstream of the TAX reduces the background
further.
A collimator at $z = 40$ m downstream of the target with $r = 16$~mm
defines the beam solid angle, and a cleaning collimator at $z = 80$~m
removes halo from particles interacting on the edges of the defining
collimator. Both are followed by horizontal sweeping magnets.
The sweeping fields have been optimized to minimize muon backgrounds.
The final stage of collimation is the active final collimator (AFC)
incorporated into the upstream veto (UV) at $z=120$~m. 
The cleaning and final (AFC) collimators have apertures which are
progressively larger than the beam acceptance, such that the former lies
outside a cone from the (2~mm radius) target passing through the defining
collimator, and the latter lies outside a cone from the TAX aperture
passing through the cleaning collimator.
The design of the beamline has been guided by FLUKA and Geant4 simulations
to quantify the extent and composition of beam halo, muon backgrounds,
and sweeping requirements.
According to the simulations, for an intensity of $2\times10^{13}$ ppp
and an effective spill length of 3~s, there are 140 MHz of $K_L$ in the beam
and 440 MHz of neutrons~\cite{vanDijk:2018xxx}.
An oriented tungsten crystal in the center of the TAX converts the vast
majority of the photons in the beam into $e^+e^-$ pairs, which are swept away,
but 40 MHz of photons with energy greater
than 5 GeV remain. As noted below, the use of an oriented
crystal converter maximizes
the transparency to neutral hadrons at a given thickness in radiation
lengths (9.4$X_0$).

\subsection{Detector}

Because of the experimental challenges involved in the measurement of
$K_L\to\pi^0\nu\bar{\nu}$, and in particular, the very high efficiency
required for the photon veto systems, 
most of the subdetector systems for KLEVER will have to be newly constructed.

\paragraph{Main electromagnetic calorimeter (MEC)}

Early studies indicated that the NA48 liquid-krypton calorimeter
(LKr) \cite{Fanti:2007vi}
currently used in NA62 could be reused as the MEC for reconstruction of
the $\pi^0$ for signal events and rejection of
events with additional photons. Indeed, the efficiency
and energy resolution of the
LKr appear to be satisfactory for KLEVER. However, the LKr time resolution
would be a major liability. The LKr would measure the event time in
KLEVER with 500 ps resolution, while the total rate of accidental vetoes
(dominated by the rate from beam photons interacting in the SAC)
could be 100 MHz.
The LKr time resolution might be improved via a comprehensive readout
upgrade, but concerns about the service life of the LKr would remain, and the
size of the inner bore would limit the beam solid angle and hence kaon flux.

We are investigating the possibility of replacing the
LKr with a shashlyk-based MEC patterned on the PANDA FS calorimeter (in turn,
based on the KOPIO calorimeter~\cite{Atoian:2007up}), with
$110\times110$~mm modules, lead absorber thickness of 0.275 mm, and scintillator
thickness of 1.5~mm, read out by silicon photomultipliers via 1-mm
diameter WLS fibers. Stochastic energy and time resolutions of better than
$\sigma_E/E = 2\%/\sqrt E$ and $\sigma_t = 72~{\rm ps}/\sqrt E$ were
obtained with this design in KOPIO tests.
We envisage a shashlyk design incorporating ``spy tiles'',
consisting of 10-mm thick scintillator bricks
incorporated into the shashlyk stack but optically isolated from it and
read out by separate WLS fibers. The spy tiles are located at key points
in the longitudinal shower development: near the front of the stack,
near shower maximum, and in the shower tail. 
This provides longitudinal sampling of the shower
development, resulting in additional information for $\gamma/n$ separation.
A first test of this concept was carried out with a prototype detector at
Protvino in April 2018.

\paragraph{Upstream veto (UV) and active final collimator (AFC)}

The upstream veto (UV) rejects $K_L\to\pi^0\pi^0$
decays in the 40~m upstream of the fiducial volume where there are
no large-angle photon vetoes.
The UV is a shashlyk calorimeter with the same basic structure as the MEC
(without the spy tiles). 

The active final collimator (AFC) is inserted into a 100-mm hole in center
of the UV. The AFC is a LYSO collar counter with angled inner surfaces to
provide the last stage of beam collimation while vetoing photons from $K_L$
that decay in transit through the collimator itself. The collar is made of
24 crystals of trapezoidal cross section, forming a detector with an
inner radius of 60 mm.
The UV and AFC are both 800 mm in depth. The maximum crystal length for
a practical AFC design is about 250 mm, so the detector consists of 3 or
4 longitudinal segments. The crystals are
read out on the back side with two avalanche photodiodes (APDs).
These devices couple well with LYSO and offer high quantum efficiency and
simple signal and HV management. Studies indicate that a light yield
in excess of 4000 p.e./MeV should be easy to achieve.

\paragraph{Large-angle photon vetoes}

Because of the boost from the high-energy beam, it is sufficient for the
large-angle photon vetoes (LAVs) to cover polar angles out to 100 mrad.
The detectors themselves must have inefficiencies of less than
a few $10^{-4}$ down to at least 100 MeV, so the current NA62 LAVs based
on the OPAL lead glass cannot be reused~\cite{Cortina-Gil:2017xxx}
The 25 new LAV detectors for KLEVER are
lead/scintillating-tile sampling calorimeters with wavelength-shifting fiber
readout, based on the CKM VVS design~\cite{Ramberg:2004en}.
Extensive experience with this
type of detector (including in prototype tests for NA62) demonstrates that
the low-energy photon detection efficiency will be sufficient for
KLEVER~\cite{Atiya:1992vh,Comfort:2005xx,Ambrosino:2007ss}.

\paragraph{Small-angle calorimeter}

The small-angle calorimeter (SAC) sits directly in the neutral beam and must
reject photons from $K_L$ decays that would otherwise escape via the
downstream beam exit. The veto efficiency required is not intrinsically
daunting (inefficiency $<1$\% for $5~{\rm GeV} < E_\gamma < 30~{\rm GeV}$
and $<10^{-4}$ for $E_\gamma > 30~{\rm GeV}$; the SAC can be blind for
$E_\gamma < 5~{\rm GeV}$),
but must be attained while maintaining insensitivity to more than 500~MHz
of neutral hadrons in the beam. In addition, the SAC must have good
longitudinal and transverse segmentation to provide $\gamma/n$
discrimination. In order to keep the false veto rate from accidental
coincidence of beam neutrons to an acceptable level ($<10$~MHz), and
assuming that about 25\% of the beam neutrons leave signals above threshold
in the SAC, topological information from the SAC must allow residual
neutron interactions to be identified with 90\% efficiency while maintaining
99\% detection efficiency for photons with $E_\gamma > 5~{\rm GeV}$. 

An intriguing possibility for the construction
of an instrument that is sensitive to photons and blind to hadrons is to
use a compact Si-W sampling calorimeter with crystalline tungsten tiles as
the absorber material, since coherent interactions of high-energy photons
with a crystal lattice can lead to increased rates of pair conversion
relative to those obtained with amorphous
materials~\cite{Bak:1988bq,Kimball:1985np,Baryshevsky:1989wm}.
The effect is dependent on photon
energy and incident angle; in the case of KLEVER, one might hope to decrease
the ratio $X_0/\lambda_{\rm int}$ by a factor of 2--3. The same effect could
be used to efficiently convert high-energy photons in the neutral beam to
$e^+e^-$ pairs at the TAX for subsequent sweeping, thereby
allowing the use of a thinner converter to minimize the scattering of
hadrons from the beam. Both concepts were tested in summer 2018 in the
SPS H2 beam line, in a joint effort together with the AXIAL collaboration.
In these tests, a beam of tagged photons with energies of up to 80 GeV was
obtained from a 120-GeV electron beam; interactions of the beam with
crystalline tungsten samples of 2 mm and 10 mm thick were studied as a
function of photon energy and angle of incidence with detectors just
downstream of the samples to measure charged multiplicity and forward
energy. While the data are still under analysis, the initial results show
a significant increase for both samples in the probability for
electromagnetic interactions when the crystal axis is aligned with the beam,
corresponding to the expected shortening of the radiation
length, with a somewhat smaller effect in the (lower-quality) thicker
crystal counterbalanced by a larger angular range over which the effect
is observed (a few mrad vs.\ about 1 mrad).

\paragraph{Charged-particle rejection}

For the rejection of charged particles, $K_{e3}$ is a benchmark channel
because of its large BR and because the final state electron can be mistaken
for a photon. Simulations indicate that the needed rejection can be achieved
with two planes of charged-particle veto (CPV) each providing 99.5\%
detection efficiency, supplemented by the $\mu^\pm$ and $\pi^\pm$ recognition
capabilities of the MEC (assumed in this case to be equal to those of the LKr)
and the current NA62 hadronic calorimeters and muon vetoes, which could be
reused in KLEVER. The CPVs are positioned $\sim$3~m upstream of the MEC
and are assumed to be constructed out of thin
scintillator tiles.
In thicker scintillation hodoscopes, the detection inefficiency arises
mainly from the gaps between scintillators. For KLEVER, the scintillators
will be only $\sim$5 mm thick (1.2\% X0), and the design will be carefully
optimized to avoid insensitive gaps.

\paragraph{Preshower detector}

The PSD measures the directions for photons incident on the MEC.
Without the PSD, the $z$-position of the $\pi^0$ decay vertex can only be
reconstructed by assuming that two clusters on the MEC are indeed
photons from the decay of a single $\pi^0$. With the PSD, a vertex can be
reconstructed by projecting the photon trajectories to the beamline.
The invariant mass is then an independent quantity, and
$K_L\to\pi^0\pi^0$ decays with mispaired photons can be efficiently
rejected.
The vertex can be reconstructed using a single photon and the constraint
from the nominal beam axis. 
Simulations show that with $0.5 X_0$ of converter
(corresponding to a probability of at least one conversion of 50\%)
and two tracking planes with a spatial resolution of 100~$\mu$m,
placed 50 cm apart, the mass resolution is about 20~MeV and the
vertex position resolution is about 10~m. The tracking detectors
must cover a surface of about 5 m$^2$ with minimal material.
Micropattern gas detector (MPGD) technology seems perfectly suited for
the PSD. Information from the PSD will be used for bifurcation
studies of the background and for the selection of control samples,
as well as in signal selection.

\subsection{Simulation and performance}

Simulations of the experiment carried out with fast-simulation techniques
(idealized geometry, parameterized detector response, etc.) show that the
target sensitivity is achievable (60 SM events with $S/B = 1$). Background
channels considered at high simulation statistics include $K_L\to\pi^0\pi^0$
(including events with
reconstructed photons from different $\pi^0$s and events with overlapping
photons on the MEC), $K_L\to 3\pi^0$, and $K_L\to\gamma\gamma$.
\begin{figure}[htb]
  \centering
  \includegraphics[width=0.8\textwidth]{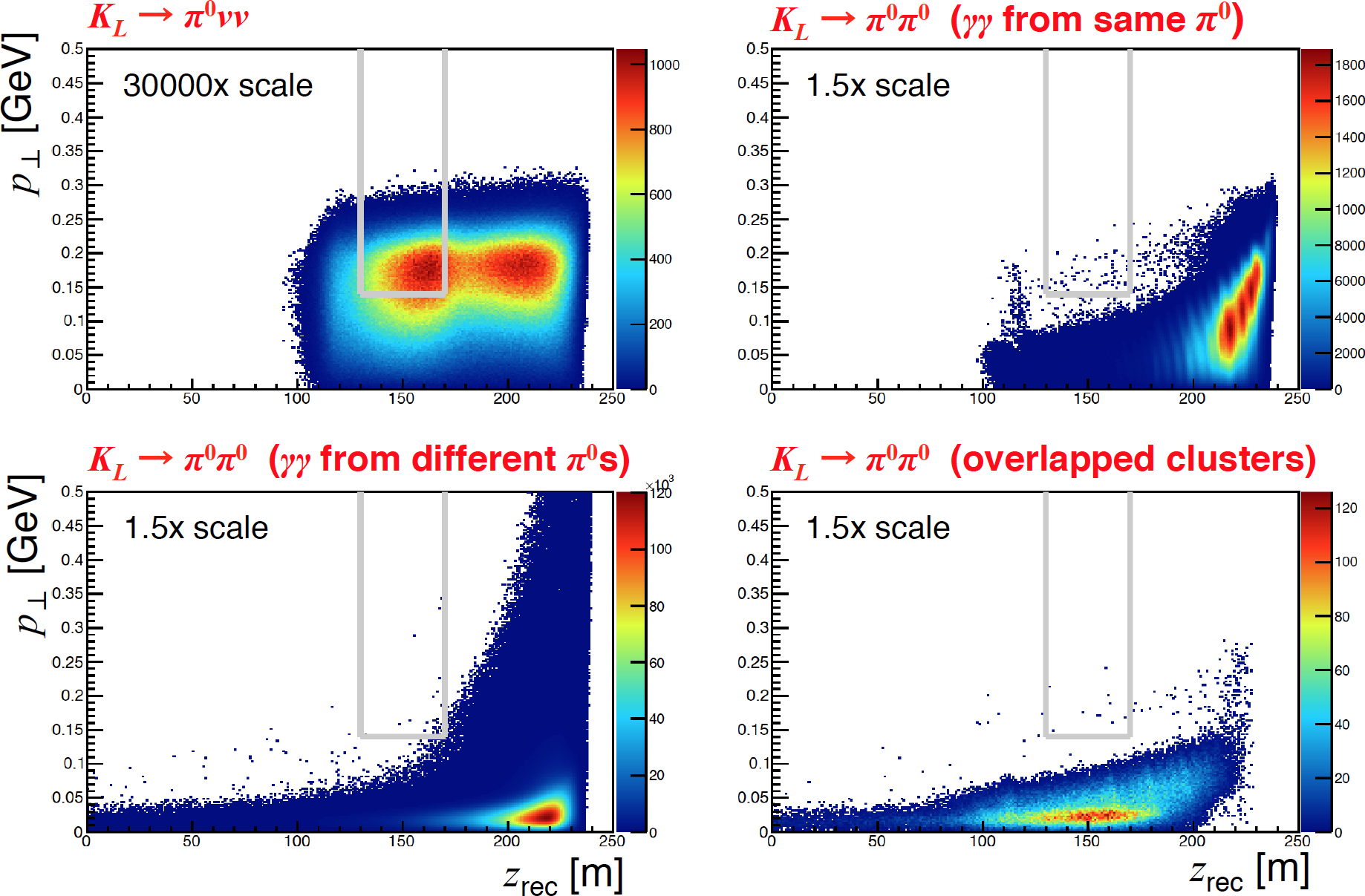}
  \caption{Distributions of events in plane of $(z_{\rm rec}, p_\perp)$
    after basic event selection cuts, from fast MC simulation, for 
    $K_L\to\pi^0\nu\bar{\nu}$ events (top left) and for
    $K_L\to\pi^0\pi^0$ events with two photons from the same
    $\pi^0$ (top right),
    two photons from different $\pi^0$s (bottom left), and
    with two or more indistinguishable overlapping photon clusters
    (bottom right).}
  \label{fig:sel}
\end{figure}
Fig.~\ref{fig:sel} illustrates the scheme for differentiating signal
events from $K_L\to\pi^0\pi^0$ background. Events with exactly two photons
on the MEC and no other activity in the detector are selected. The clusters
on the MEC from both photons must also be  more than 35~cm from the beam axis
(this helps to increase the rejection for events with overlapping clusters).
If one or both photons convert in the PSD, the reconstructed vertex
position must be inside the fiducial volume. The plots show the
distributions of the events satisfying these minimal criteria in the
plane of $p_\perp$ vs.\ $z_{\rm rec}$ for the $\pi^0$, where the
distance from the $\pi^0$ to the MEC is obtained from the transverse
separation of the two photon clusters, assuming that they come from a
$\pi^0$ ($M_{\gamma\gamma} = m_{\pi^0}$). This scheme is far from final and
there is room for improvement with a multivariate analysis, but is does
demonstrate that it should be possible to obtain $S/B\sim1$ with respect
to other $K_L$ decays.
Background
from $\Lambda\to n\pi^0$ and from decays with charged particles is assumed
to be eliminated on the basis of studies with more limited statistics.
In particular, the background from $\Lambda\to n\pi^0$ is reduced by
several orders of magnitude by the 130-m
length from the target to the fiducial volume and the choice
of production angle, which is carefully optimized to balance $K_L$ flux
against the need to keep the $\Lambda$ momentum spectrum soft.
Residual background can then be effectively eliminated
by cuts on $p_\perp$ and in the $\theta$ vs. $p$ plane for the $\pi^0$.  
The background from single $\pi^0$ production in interactions of beam
neutrons on residual gas has been estimated to be a most a few percent of
the expected signal for a residual gas pressure of $10^{-7}$ mbar.

An effort is underway to develop a comprehensive simulation
and use it to validate the results obtained so far. Of particular note,
backgrounds from radiative $K_L$ decays and cascading hyperon decays
remain to be studied, and the neutral-beam halo from our detailed
FLUKA beamline simulation needs to be fully incorporated into
the simulation of the experiment.
While mitigation of potential background contributions from one or more
of these sources might ultimately require specific modifications to the
experimental setup, we expect this task to be less complicated than
dealing with the primary challenges from $K_L\to\pi^0\pi^0$ and
$\Lambda\to n\pi^0$.

\section{Conclusions}

Significant progress on the KLEVER design has been made as a part of
and with support from the Physics Beyond Colliders initiative at
CERN in preparation for the 2020 update of the European Strategy for
Particle Physics~\cite{Ambrosino:2019xxx}.
The studies described here are preliminary.
In particular, high-statistics background studies have been carried out
for only a few channels. Some potential backgrounds, such as backgrounds
from decays with charged particles emitted in the very forward region,
remain to be investigated in detail, and some of the
detector concepts require validation.
With these caveats, our design studies indicate that an
experiment to measure ${\rm BR}(K_L\to\pi^0\nu\bar{\nu})$ can be performed
at the SPS.
An expression of interest to the SPSC is in preparation.

The KLEVER project builds on a CERN tradition of groundbreaking experiments
in kaon physics, following on the successes of NA31, NA48, NA48/2, and NA62.
Many institutes currently participating in NA62
have expressed support for and interest in the KLEVER project.
Successfully carrying out the KLEVER experimental program
will require the involvement of new institutions and groups, and we are
actively seeking to expand the collaboration at present.

\section*{References}
\bibliographystyle{iopart-num}
\bibliography{kaon19_proc}

\end{document}